\documentclass[twocolumn,showpacs,preprintnumbers,amsmath,amssymb,prb]{revtex4}
\newcommand{\be}{\begin{equation}}
\newcommand{\ee}{\end{equation}}
\newcommand{\bea}{\begin{eqnarray}}
\newcommand{\eea}{\end{eqnarray}}
\usepackage{graphicx}
\usepackage{float,afterpage}
\usepackage{dcolumn}
\usepackage{bm}
\begin{document}
\title{
A quantum mechanical relation connecting time, 
temperature, and 
cosmological constant of the universe: Gamow's relation revisited
as a special case}
%A Quantum Mechanical Derivation of Gamow's Relation for the Time and
%Temperature of the Expanding Universe: An Estimation of Cosmological 
%Constant}
\author{Subodha Mishra}
\affiliation {Department of Physics \& Astronomy,
 University of Missouri, Columbia, MO 65211, USA}
\date{\today}
%\maketitle
%\vspace{0.2 in}
%\draft
%\preprint{ }
\begin{abstract}
Considering our expanding universe as made up of gravitationally 
interacting particles which describe particles of luminous matter, 
dark 
matter and  dark energy which is described by a repulsive harmonic 
potential among the points in the flat 3-space, we derive 
a quantum mechanical relation connecting,
temperature of the cosmic microwave background radiation, age, and
cosmological constant of the universe.
When the  cosmological constant is zero, we get back Gamow's relation 
with a 
much better coefficient. Otherwise, our 
theory predicts a value of the cosmological constant $2.0\times 10^{-56}\ 
{\rm 
{cm^{-2}}}$ when 
the present values of cosmic microwave background temperature of 2.728 K 
and age of the 
universe $14$ billion years are taken as input.
\end{abstract}
\maketitle

%\vskip5mm\noindent

\section{Introduction}
The most important theory for the
origin of the universe is the Big Bang Theory\cite{alan} according to 
which
the present universe is considered to have started with a huge 
explosion from a superhot and a superdense stage. Theoretically
one may visualize its starting from a mathematical singularity with
infinite density. This also comes from the solutions of the type I and type II
form of Einstein's field equations\cite{cont}. What follows from all these 
solutions is 
that the universe has originated from a point where the scale factor $R$ (to be 
identified as the radius of the universe) is zero at time $t = 0$,
and its derivative with time is taken to be infinite at this time. That is, it is 
thought that the initial explosion had happened with infinite velocity,
although, it is impossible for us to picture the initial moment of 
the creation of the universe. The accelerated expansion of the universe 
has been conformed by studying the distances to supernovae of 
type Ia\cite{riess,per}.
For the universe, it is being said
that the major constituent of the total mass of the present universe
is made of the Dark Energy 70$\%$, Dark Matter about 26$\%$ and luminous 
matter $4\%$. The Dark energy is responsible for the accelerated 
expansion of the universe since it has negative pressure and 
produces repulsive gravity. The cosmological constant\cite{peb,har} of 
Einstein 
provides a repulsive force when its value  is  positive.
The cosmological constant is also associated with the vacuum energy 
density\cite{sean} of the space-time. The vacuum has the lowest 
energy of any state, but there is no reason in principle for that 
ground state energy to be zero. There are 
many different 
contributions\cite{sean} to the ground state energy such as potential 
energy of  scalar fields, vacuum 
fluctuations as well as of the cosmological 
constant. The 
individual contributions can be very large but current observation 
suggests that the various contributions, large in magnitude but 
different in sign delicately cancel to yield an extraordinarily small 
final result. The conventionally defined cosmological constant $\Lambda$ 
is proportional to the vacuum energy density $\rho_\Lambda$ as 
$\Lambda=(8\pi G/c^2)\rho_\Lambda$. Hence one can guess that 
$\rho_\Lambda=\Lambda c^2/8\pi G\approx\rho_{Pl}=c^5/G^2\hbar\sim 
5\times 10^{93}\ g \ cm^{-3}$, where $\rho_{Pl}$ is the Plank density. 
But the recent observations of the luminosities of high redshift 
supernovae gives the dimensionless density 
$\Omega_\Lambda=\rho_\Lambda/\rho_{cr}\equiv\Lambda c^2/3H_0^2\approx 
0.7$ where $\rho_{cr}=3H_0^2/8\pi G\approx 1.9\times 10^{-29}\ g\ 
cm^{-3}$, 
which implies $\rho_\Lambda =\rho_{Pl}\times 10^{-123}$. This shows that 
the cosmological constant today is 123 orders of magnitude 
smaller. 
%Or one can say that one can get a dimensionless combination 
%$\Lambda (G\hbar/c^3)\equiv\Lambda\L^2_P$.
% But the bound from 
%obesrevation gives us $\Lambda L^2_P\le 10^{-123}$. 
This is known as the  
'cosmological constant problem'.  
In the classical big-bang cosmology there is no dynamical theory\cite{nar1} 
to relate 
the cosmological constant to any other physical variable of the universe.
There have been some studies\cite{vol,laugh,hu} regarding the universe to 
relate the space-time manifold to somekind of condensed matter 
systems.
Here by considering\cite{dn,dnt} the visible universe made up of 
self-gravitating 
particles 
representing luminous baryons and dark matter such as neutrinos ( though 
only a small fraction) which are fermions and a repulsive potential 
describing the effect of Dark Energy  responsible for the 
accelerated expansion 
of the universe, we in this
 paper derive quantum mechanically a relation connecting temperature, 
age and cosmological constant of the universe. When the  cosmological 
constant is zero, we get back Gamow's relation with a much better 
coefficient. 
Otherwise using as input the current
values of $T=2.728\ K \ {\rm and}\  t=14\times 10^9\ years$,
we predict the value of cosmological constant as $2.0\times 10^{-56}\  
cm^{-2}$. 
Note 
that 
$\Lambda$ is a 
completely free parameter in General Theory of Relativity. Also it is 
interesting to note that we obtain not only the value of the 
cosmological constant but also the sign of the parameter correct 
though it is a very small number.

\section{Mathematical formulation without the cosmological constant 
$\Lambda$}
We in this section derive a relation connecting temperature and age of the 
universe when cosmological constant is zero, by considering a 
Hamiltonian\cite{dn,dnt,sm} 
used by us some time back  for the study of a system of
self-gravitating particles which is given as:  
\be
H = - \sum^N_{i=1}({\hbar^2\over 2m})\nabla_i^2+{1\over 2}\sum^N_{i=1,}
\sum^N_{i\not =j,j=1} v (\mid\vec X_i-\vec X_j\mid)\label{equh}
\ee
where $v(\mid\vec X_i-\vec X_j\mid) = - g^2/\mid\vec X_i-\vec
X_j\mid$, having $g^2=Gm^2$, 
$G$ being the universal gravitational constant and $m$ the mass of the
effective constituent particles describing the luminous matter and 
dark matter whose number is $N=\int \rho(\vec X) d\vec X$. 
Since the measured value for the temperature of the cosmic microwave
background radiation is $\approx 2.728 K$, it lies in the neighbourhood
of almost zero temperature. We,therefore, 
use the zero temperature formalism for the study of the present problem.
Under the situation $N$ is
extremely large, the total kinetic energy of the system is obtained
as 
\be
<KE> = \bigg ({3\hbar^2\over 10m}\bigg )(3 \pi^2)^{2/3}\int d\vec X
[\rho(\vec X)]^{5/3}\label{ke1}
\ee
where $\rho(\vec X)$ denotes the single particle density to account
for the distribution of particles (fermions) within the system, which is
considered to be a finite one. Eq.(\ref{ke1}) has been written in the
Thomas-Fermi approximation. The total potential energy of the system,
in the Hartree-approximation, is now given as
\be
<PE> = - ({g^2\over 2})\int d\vec X d\vec X'{1\over \mid\vec
X-\vec X' \mid} \rho(\vec X)\rho(\vec X')\label{pe1}
\ee
Inorder to evaluate the integral in Eq.(\ref{ke1}) and Eq.({\ref{pe1}), we had
chosen a trial single-particle density\cite{dn,dnt} $\rho(\vec X)$ 
which was of
the form :
\be
\rho(\vec X) = {{A e^{-x}}\over x^3}\label{den}
\ee
where $x = (r/\lambda)^{1/2}$, $\lambda$ being the variational
parameter. As one can see from Eq.(\ref{den}), $\rho(\vec X)$ is singular at
the origin.
This looks to be consistent with the concept behind the Big Bang theory of 
the universe. The early universe was not only known to be super hot, but 
also it was superdense. To account for the  scenario at the time of the 
Big Bang, we have, therefore, imagined of a single-particle density 
$\rho(x)$ for the system which is singular at the origin $(r=0)$. 
This is only true at the microscopic level, which is not so 
meaningful looking at things macroscopically. Although $\rho(x)$ is 
singular, the number of particles N, in the 
system is finite. 
Since the present universe has a finite size, its present density which is 
nothing but an average value is finite.
At 
the time of Big Bang $(t=0)$, since the scale factor (identified as the 
radius of the universe) is 
supposed to be  zero, the average density of the system can assume an 
infinitely large value, implying its superdense state. Having thought of a 
singular form of single-particle density at the  time of the Big Bang, we 
have tried with a number of singular form of single particle densities of 
the kind 
$\rho(\vec r)=A\frac{exp[-(\frac{r}{\lambda})^{\nu}]}
{(\frac{r}{\lambda})^{3\nu}}$
where $\nu=1,2,3,4...or\ \frac{1}{2},\frac{1}{3},\frac{1}{4},...$. Integer
values of $\nu$ are not permissible because they make the normalization
constant infinite. Out of the fractional values, $\nu=\frac{1}{2}$ is
found to be most appropriate, because, it has been shown in our earlier 
paper that it gives
the expected upper limit for the critical mass of a neutron
star\cite{dn} beyond which black hole 
formation takes place and other parameters of the universe\cite{dnt} 
satisfactorily correct.
Also because if $\nu$ goes to zero (like $1/n$, $n\rightarrow\infty$),
$\rho(r)$ would tend to the case of a constant density as found in an
infinite many-fermion system.
In view of the arguments put forth above, one will have to think that the 
very choice of our $\rho(r)$ is a kind of ansatz in our theory, which is 
equivalent to the choice of a trial wave function used in the quantum 
mechanical calculation for the binding energy of a physical system 
following variational techniques. As mentioned earlier, singularity at 
$r=0$ in the single particle distribution has nothing to do with the 
average particle density in the system, which happens to be finite ( 
because of the fact that N is finite and volume V of the visible universe 
is 
finite), and hence it is not going to affect the large scale spatial 
homogeneity of the observed universe.
 Having accepted the value $\nu=\frac{1}{2}$,
the parameter $\lambda$ associated with $\rho(r)$ is determined after
minimizing $E(\lambda)=<H>$ with respect to $\lambda$. This is how, we are
able to find the total energy of the system corresponding to its lowest
energy state.

 After evaluating the integrals in Eq.(\ref{ke1}) and Eq.(\ref{pe1}), we 
find 
the total energy $E(\lambda)$ of the system which is given as 
\be
E(\lambda)=\frac{\hbar^2}{m}\frac{12}{25\pi}(\frac{3\pi
N}{16})^{5/3}\frac{1}{\lambda^2}-\frac{g^2N^2}{16}\frac{1}{\lambda}\label{bige}
\ee
We minimize the total energy with respect to $\lambda$.
Differentiating this with respect to $\lambda$ and then equating it with
zero, we obtain the value of $\lambda$  at which the minimum occurs. This
is found as:
\be
\lambda_0=\frac{72}{25}\frac{\hbar^2}{mg^2}(\frac{3\pi}{16})^{2/3}
\frac{1}{N^{1/3}}\label{lam}
\ee
 Following the expression for $<KE>$
evaluated at $\lambda = \lambda_0$, we write down
the value of the equivalent temperature $T$ of the system, using the
relation
\be
T = {\frac{2}{3k_B}} [ {< KE >\over N}]\\
=  {\frac{2}{3k_B}} (0.015442) N^{4/3} ({mg^4\over
\hbar ^2})\label{t1}
\ee
The expression for the radius $R_0$ of the universe, as found by us
earlier\cite{dnt}, is given as
\be
R_0=2\lambda_0= 4.047528 ({\hbar^2\over mg^2})/N^{1/3}\label{r01}
\ee
Our identification of the radius $R_0$ with $2\lambda_0$ is
based on the use of socalled quantum mechanical tunneling\cite{karp}
effect. Classically, it is well known that a particle has a turning
point
where the potential energy becomes equal to the total energy. Since the
kinetic energy and therefore the velocity are equal to zero at such a
point, the classical particle is expected to be turned around or reflected
by the potential barrier. From the present theory it is seen that the
turning point occurs at a distance $R=2\lambda_0$.

After invoking Mach's principle\cite{har}, which is expressed through the
relation $({GM\over R_0 c^2})\approx 1$,  and using the fact that the
total mass of the universe
$M= Nm$, we are able to obtain the total number of particles $N$ constituting the
universe, as
\be
N = 2.8535954 ({\hbar c\over Gm^2})^{3/2}\label{n1}
\ee
Now, substituting Eq.(\ref{n1}) in Eq.(\ref{r01}), we arrive at the expression for
$R_0$, as
\be
R_0 = 2.8535954 ({\hbar\over mc})({\hbar c\over Gm^2})^{1/2}\label{r02}
\ee
As one can see from above, $R_0$ is of a form which involves only the fundamental constants like $\hbar, c, G$ and
$m$. Now, eliminating $N$ from Eq.(\ref{t1}), by virtue of
Eq.(\ref{n1}),we have
\be
T = {2\over 3} (0.0625019)({mc^2\over k_{B}})\label{tem1}
\ee

Let us now assume that the radius $R_0$ of the universe is
approximately given by the relation
\be
R_0 \simeq ct\label{r0ct}
\ee
where $t$ denotes the age of the universe at any instant of time.
%The Hubble's law as indicated in Eq.(\ref{hlaw}), also implies that the 
%universe
%is expanding uniformly. Although, it is so for the universe, all the
%galaxies are not uniformly expanding. Considering a photon of light with 
%wave length
%$\lambda$ travelling a distance of separation 'd' between two
%galaxies at rest with respect to
%each other, one has $d=ct$, where 't' is the time it takes for light to
%travell the space between the galaxies. Because of the expansion of
%the universe, the galaxies move away from each other at a velocity
%$v$ known as the radial velocity. During this time 't', the galaxies
%are separated by a distance $\Delta d$ given by $\Delta d = vt$.
%Thus, one finds that ${\Delta d \over d} =({v\over c})$. From this, it 
%follows that the
%greater is the relative velocities of the galaxies, the greater is
%the separation attained in the time interval $t$. The
%importance of Hubble law, as stated through Eq.(\ref{hlaw}), is that the 
%galaxies were closer in the past
%than they are now. As we have stated earlier, the Hubble time $({1\over 
%H_0})$ represents the
%maximum age of the universe, because the galaxies themselves slow
%down the expansion of the universe. Even though the galaxies are
%farther apart, they still exert a gravitational force on each other.
%Their mutual gravity continuously acts to pull other galaxies together. 
%This
%means that the universe was expanding faster in the past than it is
%now. As indicated in Eq.(\ref{r0ct}), the velocity of expansion of the 
%universe is being approximated
%to be equal to the velocity of light 'c'.
Following
Eq.(\ref{r02}) and Eq.(\ref{r0ct}), we write $m$ as
\be
m = ({\hbar^3\over Gc^3})^{1/4} (2.8535954)^{1/2} {1\over\sqrt 
t}\label{mmm}
\ee
%The variation of mass m with time is not incompatible with the fact that 
%our 
%theory is not relativistically invariant and recently a variation of 
%ratio 
%of mass of proton to mass of electron has been observed\cite{rein} which 
%shows that 
%mass of elementary particle has changed with time since the Big Bang.
It is interesting to see (as shown in Table-1) this 
variation of mass with time 
gives approximately the energy and hence the temperature scale of 
formation of elementary 
particles in 
different epochs of 
nucleosynthesis. We calculate temperatures in different epochs using our 
Eq.(\ref{tem3}) to be derived shortly. This is in  good agreement 
with the calculated 
values of  temperature otherwise known from nucleosynthesis 
calculations\cite{cont,nar1}. 
The period between $t=7\times 10^{-5}\ sec\ and \ 5\ sec$ is called lepton 
era, while period before $t=7\times 10^{-5}\sec$ is hadron era and the 
early era corresponding to the period $t<10^{-43}\ sec$ is known as Planck 
era.

%\vfill
\begin{widetext}
%\vfill
%\eject
%\magnification=\magstep1
%\hsize 30 pc
%\vsize 42 pc
\centerline {\bf {TABLE - 1 }}
%\midinsert
\vbox{\offinterlineskip
\halign{&\vrule#&\strut\ #\ \cr
\noalign{\hrule}
&\hfil Age of the   \hfil&&\hfil Temperature (T) in K as  
\hfil&&\hfil
Temperature (T)in K    &\hfil\cr
&\hfil universe (t)    \hfil&&\hfil  calculated from   \hfil&&\hfil for 
the formation of  &\hfil\cr
&\hfil in sec. \hfil&&\hfil  Eq. (15)   \hfil&&\hfil elementary 
particles\cite{cont,nar1}    
&\hfil\cr
height3pt&\omit&&\omit&&\omit&\cr
\noalign{\hrule}
&\hfil 5   \hfil&&\hfil $\approx 1\times 10^9$   \hfil&&\hfil $\approx 
6\times 10^9 (e^+,e^-)$    &\hfil\cr
height3pt&\omit&&\omit&&\omit&\cr
&\hfil $1.2\times 10^{-4}$   \hfil&&\hfil $\approx 2.1\times 10^{11}$
\hfil&&\hfil $\approx 1.2\times 10^{12} (\mu^+,\mu^-$ and their 
antiparticles)   &\hfil\cr
height3pt&\omit&&\omit&&\omit&\cr
&\hfil $7\times 10^{-5}$   \hfil&&\hfil $\approx 2.8\times 10^{11}$
\hfil&&\hfil $\approx 1.6\times 10^{12} (\pi^0,\pi^+,\pi^-$ and
their antiparticles)    &\hfil\cr
height3pt&\omit&&\omit&&\omit&\cr
&\hfil $1.5\times 10^{-6}$   \hfil&&\hfil $\approx 1.9\times 10^{12}$
\hfil&&\hfil $\approx  10^{13}$ (protons, neutron and their antiparticles)   
&\hfil\cr
height3pt&\omit&&\omit&&\omit&\cr
&\hfil $10^{-43}$   \hfil&&\hfil $\approx 0.73\times 10^{31}$
\hfil&&\hfil $\approx  10^{32}$ (planck \ mass)   &\hfil\cr
height3pt&\omit&&\omit&&\omit&\cr
height3pt&\omit&&\omit&&\omit&\cr
\noalign{\hrule}
\noalign{\hrule}\noalign{\smallskip}}}
%$$\endinsert
%\vfill
%\eject
\end{widetext}
A substitution of $m$, from Eq.(\ref{mmm}), in Eq.(\ref{tem1}),
 enables us to write
\bea
T &=& 0.070388 ({1\over k_{B}}) ({c^5\hbar^3\over G})^{1/4} t^{-1/2}
\nonumber \\
&=& 0.06339 [\frac{c^2}{G\ a_B}]^{1/4}
t^{-1/2}\label{tem2}
\eea
This is exactly the Gamow's relation\cite{nar1,sm} apart from the fact
that
Gamow's
relation had the coefficient 0.41563 instead of 0.06339 as in our
expression.
Substituting the numerical value of $a_B$, which is equal to $7.56\times
10^{-15} \ erg\  cm^{-3} K^{-4} \
$, and the present value for the universal gravitational constant
$G$ $[G = 6.67 \times 10^{-8} dyn. cm^2. gm^{-2}]$, in Eq.(\ref{tem2}),we 
obtain
\be
T = (0.23172 \times 10^{10}) t_{sec}^{-1/2} K\label{tem3}
\ee
%As one can very well see, Eq.(\ref{tem3}) is of the same form as obtained
%by
%Gamow [Eq.(\ref{gae})] leaving aside the
%multiplying constant 0.23172.
If we accept the age of the universe to
be close to $14\times 10^9 year$, which we have used here, with the
help of Eq.(\ref{tem3}), we arrive at a value for
 the Cosmic Microwave Background Temperature (CMBT)
equal to $\approx 3.5 K$. This is very close to the measured value
of 2.728 K as reported from the most recent Cosmic Background
Explorer (COBE) satellite measurements\cite{penz,kobe}. However, if we use
Gamow's
relation,  $t=956$ billion years is required
to
obtain the exact value of 2.728 K for the cosmic background temperature
from. Using our
expression, Eq.(\ref{tem3}), we would require an age of $22.832 \times
10^9\ year$ for the universe to get the exact value of 2.728 K. Long back 
a
correction was made  to Gamow's relation by multiplying it with a factor 
of
$(\frac{2}{g_d})^{1/4}$ by taking into account the degeneracies of the
particles, where $g_d=9$. This correction effectively multiplies
Gamow'relation with a
factor of 0.68 and brings back the  age of the universe to 425 billion
years for the present CMBT. If we 
multiply our 
expression by the same
factor to correct for the degeneracy of particles, we obtain a value of
2.4 K, which is less than the value of
present CMBT. In the next section we see that by including the cosmic
repulsion by the part
given by cosmological constant we get back 2.728 K, This is physically
correct since the cosmological term\cite{har} has the meaning of negative
pressure,
it adds energy to the system by its tension when the universe expands,
though the over all temperature decreases as the universe expands.

\section{Inclusion of the part of the Hamiltonian corresponding to the 
Cosmological constant}

The cosmological constant term\cite{peb,har} $\Lambda$ associated with 
vacuum 
energy density
 was originally 
introduced 
by Einstein as a repulsive component in his field equation and 
when 
translated from  the relativistic to Newtonian picture gives rise to a 
repulsive harmonic oscillator force per unit mass as $\sim (\Lambda 
c^2)\vec 
r$ between points in space 
when $\Lambda$ 
is positive. The one-body operator corresponding to the potential can 
be written as     
$H_\Lambda=-{{\Lambda}}c^2|\vec X|^2\rho(\vec 
X)$ where the density term takes care of the unit mass 
in the repulsive potential. Hence
the energy corresponding to this repulsive potential 
can be written as: 
%\be
%H_\Lambda=-\frac{1}{2}\sum^N_{i=1}\sum^N_{j=1}{\frac{\Lambda}{3}} 
%c^2|\vec X_i-\vec X_j|^2
%\ee
\be
<H_\Lambda>=-\int{{\Lambda}}
c^2|\vec X|^2\rho^2(\vec X)\ d\vec X
\ee
By including this contribution of $H_{\Lambda}$ in Eq.(\ref{bige}), we 
have 
the total energy
\be
E(\lambda)=\frac{\hbar^2}{m}\frac{12}{25\pi}(\frac{3\pi
N}{16})^{5/3}\frac{1}{\lambda^2}-\frac{g_{\Lambda}^2N^2}{16}.\frac{1}{\lambda}
\ee
where $g_{\Lambda}^2=g^2+\frac{3\Lambda c^2}{16\pi}$.
Calculating as before, we have
\be
N = 2.8535954 ({1\over Gm^2})^{3/4}({\hbar c\over 
g_{\Lambda}})^{3/2}\label{n1l}
\ee
and
\be
R_0 = 4.047528 ({\hbar\over mc})^{1/2}({\hbar G^{1/4}\over 
g_\Lambda^{3/2}})\label{r02l}
\ee
Now equating this $R_0$ with $ct$ we have
\be
Gm^{8/3}-\frac{3\Lambda c^2}{16\pi} m^{2/3}-Q=0
\ee
where
$Q=\frac{4.0475279 
\hbar^{2}G^{1/3}}{c^{2}}\times \frac{1}{t^{4/3}}$.
Using $m'=m^{2/3}$, the above equation can be cast as a quartic 
equation in  $m'$. We find\cite{ab} four
analytic solutions for $m'$ and hence for $m$. Three of the solutions are 
unphysical and the only solution which is physically correct is given as

%\be
%m=(\frac{-u^{1/2}\pm\sqrt{u-4(u/2+[(u/2)^2+Q/G]^{1/2})}}{2})^{3/2}
%\ee
%and
\be
m=(\frac{u^{1/2}+\sqrt{u-4(u/2-[(u/2)^2+Q/G]^{1/2})}}{2})^{3/2}\label{mas}
\ee
where
\be
u=[r+(q^3+r^2)^{1/2}]^{1/3}\ +\ [r-(q^3+r^2)^{1/2}]^{1/3}
\ee
and $r=\frac{9\Lambda^2 c^4}{2(16\pi G)^2}$,\ $q=\frac{4Q}{3G}$.
Now the Kinetic energy with the degeneracy factor as discussed in the 
previous section, is given as
\be
T = ({\frac{2}{g_d}})^{\frac{1}{4}}{\frac{2}{3k_B}} [ {< 
KE >\over N}]\\
=({\frac{2}{g_d}})^{\frac{1}{4}}{\frac{2}{3k_B}} 
(0.015442) N^{4/3} 
({mg_\Lambda^4\over\hbar^2})\label{kel}
\ee
Using Eq.({\ref{n1l}) and Eq.({\ref{mas}}) in Eq.({\ref{kel}}), we finally 
have the  relation,
\begin{widetext}
\be
T=0.0417({\frac{2}{g_d}})^{1/4}{\frac{c^2}{k_B}}
\frac{[(\{{u^{1/2}+\sqrt{4[(u/2)^2+Q/G]^{1/2}-u}}\}/{2})^{3}
+{\frac{3\Lambda c^2}{16\pi G}}]}
{(\{{u^{1/2}+\sqrt{4[(u/2)^2+Q/G]^{1/2}-u}}\}/{2})^{3/2}}\label{gt}
\ee
\end{widetext}
%{(\frac{u^{1/2}\pm\sqrt{u-4(u/2-[(u/2)^2+Q/G]^{1/2})}}{2})^{3/2}}\label{t1l}
%\end{widetext}
This is the central result of our paper.
This relation connects temperature T with time $t$ and cosmological 
constant $\Lambda$ since Q is a function of $t$ and u is also a function 
of $t$ and $\Lambda$. When $\Lambda$=0, we get back 
the 
relation Eq.(\ref{tem2}) connecting T and t. Since we know the current 
values of $T=2.728K \ {\rm and}\  t=14\times 10^9 year$, using 
Eq.(\ref{gt}), 
we 
solve for $\Lambda$. We do that in Fig.\ref{fig1} by plotting the left 
hand side 
and right 
hand side of Eq. (\ref{gt}) and finding the crossing point. This gives 
$\Lambda=2.0\times 10^{-56}\ 
{\rm cm^{-2}}$ 
which is the value that has been 
derived dynamically here.

\begin{figure}[htb]
\includegraphics[angle=0,width=0.75\linewidth]
{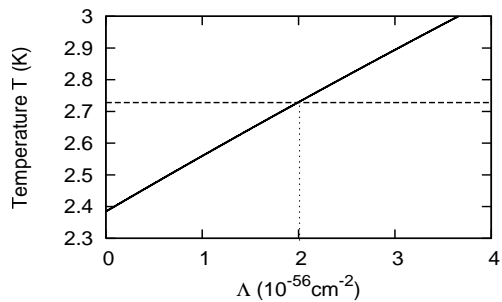}
\caption{\label{templ}
Determination of $\Lambda$ by plotting the right hand side of 
Eq.(\ref{gt}) 
as a function of $\Lambda$ (solid line) and left hand side as 2.728 K 
(thin broken line). 
The vertical 
dotted line indicates the value of $\Lambda=2.0\ 10^{-56}cm^{-2}$
}\label{fig1}
\end{figure}

\section{Conclusion}

To conclude, we in this letter have derived a relation connecting 
temperature, age and cosmological constant of the universe by describing 
the  universe as made up of self-gravitating particles effectively 
representing luminous matter, dark matter and dark energy 
represented by the repulsive potential given by the cosmological constant.  
When the cosmological constant is zero, we get back Gamow's relation with 
a better coefficient. Other wise our theory predicts the value of 
cosmological constant as $2.0\times 10^{-56}\ {\rm cm^{-2}}$. It is 
interesting 
to note that in this flat universe, our method  
dynamically determines the value of the cosmological constant reasonably 
well compared to General Theory of Relativity where the  cosmological 
constant is a free parameter.

We thank B. Mashhoon for his useful comments on the manuscript.

%\vfill
%\eject
%\newpage
%.
%\newpage
%.

\end{document}